\documentclass{my-ws-p10x7}

\begin{document}
\def\be{\begin{eqnarray}}
\def\en{\end{eqnarray}}
\def\non{\nonumber}
\def\la{\langle}
\def\ra{\rangle}
\def\nc{N_c^{\rm eff}}
\def\lsim{ {\ \lower-1.2pt\vbox{\hbox{\rlap{$<$}\lower5pt\vbox{\hbox{$\sim$}
}}}\ } }
\def\gsim{ {\ \lower-1.2pt\vbox{\hbox{\rlap{$>$}\lower5pt\vbox{\hbox{$\sim$}
}}}\ } }
\def\pr{{\it Phys. Rev.}~}
\def\prl{{\it Phys. Rev. Lett.}~}
\def\pl{{\it Phys. Lett.}~}
\def\np{{\it Nucl. Phys.}~}
\def\zp{{\it Z. Phys.}~}

\title{Implications of Recent Measurements
of Nonleptonic Charmless $B$ Decays}

\author{Hai-Yang Cheng}

\address{Institute of Physics, Academia Sinica, Taipei, Taiwan 115, R.O.C.\\E-mail:
phcheng@ccvax.sinica.edu.tw}

\twocolumn[\maketitle\abstract{\centerline{Implications of recent
measurements of hadronic charmless $B$ decays are discussed. }}]
\section{$B\to \pi\pi$, $\pi K$ Decays}
Both $B$ factories BELLE \cite{Belle} and BABAR \cite{BaBar} have
reported at this conference the preliminary results of $B\to K\pi$
and $B\to\pi^+\pi^-$ (see Table I).  For the unitary angle
$\gamma\sim 60^\circ$, the ratio $R={\cal B}(B\to
K^\pm\pi^\mp)/{\cal B}(B\to\pi^+\pi^-)$ is conventionally
predicted to lie in the region 1.3$-$2.0, to be compared with the
experimental values of $4.0\pm 1.6$, $2.8\pm 2.0$ and $1.3\pm 0.5$
obtained by CLEO \cite{CLEO}, BELLE and BABAR, respectively.
Hence, the expected ratio $R$ is smaller than the CLEO and BELLE
results and in agreement with BABAR. Of course, more data are
needed to clarify the issue.

\begin{table*}[t]
\caption{Experimental values of the branching ratios (in units of
$10^{-6}$) for $B\to K\pi$ and $\pi\pi$.}
\begin{center}
\begin{tabular}{|c|c c p{58pt}|}
\hline \raisebox{0pt}[10pt][6pt]{Decay} & CLEO \cite{Cronin,CLEO}
& BELLE \cite{Belle} & BABAR \cite{BaBar}
 \\[3pt] \hline
\raisebox{0pt}[10pt][6pt]{$K^\pm\pi^\mp$}  &
$17.2^{+2.5}_{-2.4}\pm1.2$ & $17.4^{+5.1}_{-4.6}\pm3.4$ &
$12.5^{+3.0+1.3}_{-2.6-1.7}$ \\[6pt]
\raisebox{0pt}[10pt][6pt]{$K^0\pi^\pm$}  &
$18.2^{+4.6}_{-4.0}\pm1.6$ & $16.6^{+9.8+2.2}_{-7.8-2.4}$ & {}
\\[6pt] \raisebox{0pt}[10pt][6pt]{$K^\pm\pi^0$}  &
$11.6^{+3.0+1.4}_{-2.7-1.3}$ & $18.8^{+5.5}_{-4.9}\pm2.3$ & {}
\\[6pt] \raisebox{0pt}[10pt][6pt]{$K^0\pi^0$}  &
$14.6^{+5.9+2.4}_{-5.1-3.3}$ & $21.0^{+9.3+2.5}_{-7.8-2.3}$ & {}
\\[6pt] \hline \raisebox{0pt}[12pt][6pt]{$\pi^\pm\pi^\mp$}  &
$4.3^{+1.6}_{-1.4}\pm0.5$ & $6.3^{+3.9}_{-3.5}\pm1.6$ &
$9.3^{+2.6+1.2}_{-2.3-1.4}$ \\[6pt]
\raisebox{0pt}[8pt][6pt]{$\pi^\pm\pi^0$}  & $<12.7$ & $<10.1$ &
 \\[6pt] \raisebox{0pt}[5pt][6pt]{$\pi^0\pi^0$}  & $<5.7$ &  &
 \\[6pt]
\hline

\end{tabular}
\end{center}
\end{table*}

If the CLEO or BELLE data are taken seriously, we may ask the
question of how to accommodate the data of $K\pi$ and $\pi\pi$
simultaneously. A fit to $\pi^+\pi^-$ yields $F_0^{B\pi}(0)<0.25$
for $|V_{ub}/V_{cb}|=0.09$ and $\gamma=60^\circ$. The $K\pi$ rates
will then become too small compared to the data. By contrast, a
fit to $K\pi$ modes usually implies a too large $\pi^+\pi^-$ rate.
There are several possibilities that the CLEO or BELLE data of
$K^\pm\pi^\mp$ and $\pi^+\pi^-$ can be accommodated: (1)
$\gamma\sim 60^\circ$ and $F_0^{B\pi}(0)<0.25$ with a smaller
strange quark mass, say $m_s(m_b)=60$ MeV. The idea is that the
$K\pi$ mode will receive a sizable $(S-P)(S+P)$ penguin
contributions, while the $\pi\pi$ decay is not much affected.
However, a rather smaller $m_s$ is not consistent with recent
lattice calculations. (2) $\gamma\sim 60^\circ$ and
$F_0^{B\pi}(0)=0.30$ with the following cases: (i) a smaller
$V_{ub}$, say $|V_{ub}/V_{cb}|\approx 0.06$, so that the
$\pi^+\pi^-$ rate is suppressed. However, this small $V_{ub}$ is
not favored by data. (ii) a large nonzero isospin $\pi\pi$ phase
shift difference of order, say $70^\circ$, can yield a substantial
suppression of the $\pi^+\pi^-$ mode \cite{CCTY}. However,
$\pi^0\pi^0$ will be substantially enhanced by the same strong
phase. The CLEO new measurement \cite{CLEO} ${\cal
B}(B\to\pi^0\pi^0)<5.7\times 10^{-6}$ indicates that the strong
phase cannot be too large. (iii) a large inelasticity for
$\pi^+\pi^-$ and $D^+D^-$ modes so that the former is suppressed
whereas the latter is enhanced. (3) a large $\gamma$, say
$\gamma\sim (110-130)^\circ$, and $F_0^{B\pi}(0)=0.30$. Several
calculations \cite{Du} based on generalized or QCD improved
factorization imply $\gamma>100^\circ$. This scenario is
interesting and popular, but it encounters two problems: (i) It is
in conflict with the unitary angle $\gamma=(58.5\pm7.1)^\circ$
extracted from the global CKM fit\cite{Stocchi}. (ii) The CLEO
data other than $K\pi$ and $\pi\pi$ do not strongly support a
large $\gamma$ (see below). (4) $\gamma\sim 90^\circ$ and
$F_0^{B\pi}(0)=0.25$ as assumed in a recent PQCD analysis
\cite{Keum}. In this work, the $\pi^+\pi^-$ rate is small because
of the small form factor $F_0^{B\pi}(0)$, and $K\pi$ rates are
enhanced by large penguin effects owing to steep $\mu$ dependence
of the leading-order penguin coefficients $c_4(\mu)$ and
$c_6(\mu)$ at the hard scale $t<m_b/2$.

As shown in \cite{Beneke}, the nonfactorized term is dominated by
hard gluon exchange in the heavy quark limit as soft gluon
contributions to $\chi_i$ are suppressed by orders of
$\Lambda_{\rm QCD}/m_b$. However, there is an additional chirally
enhanced corrections to the spectator-interaction diagram, which
are logarithmically divergent \cite{Beneke1}. For example, an
additional $(V-A)(V-A)$ spectator interaction proportional to the
twist-3 wave function $\phi_\sigma^\pi$ will contribute to
$B\to\pi\pi,~\pi K$. Consequently, the nonfactorized contribution
to the coefficient $a_2(\pi\pi)$, for example, can be large
\cite{CY1}; its real part lies in the range 0.17--0.25. This will
affect the prediction of $B\to\pi^+\pi^0$ and in particular
$B\to\pi^0\pi^0$. We find that even in the leading-twist
approximation, the same logarithmically divergent integral also
appears in the charm quark mass corrections to the
spectator-interaction diagram in $B\to J/\psi K(K^*)$ decays
\cite{CY1}. As a result, Re$\,a_2(J/\psi K)$ is in the vicinity of
0.22.

\section{$B\to \rho\pi,~\omega\pi$ Decays}
The class-III decays $B^\pm\to\rho^0\pi^\pm,~\omega\pi^\pm$ are
tree-dominated and sensitive to $(N_c^{\rm eff})_2$ appearing in
$a_2$; their branching ratios decrease with $(N_c^{\rm eff})_2$.
The present data\cite{Jessop}
\be
{\cal B}(B^\pm\to \rho^0\pi^\pm) &=&
(10.4^{+3.3}_{-3.4}\pm2.1)\times 10^{-6}, \non\\ {\cal B}(B^\pm\to
\omega\pi^\pm) &=& (11.3^{+3.3}_{-2.9}\pm 1.5 )\times 10^{-6},
\non \\
\en
imply $(\nc)_2<3$ as in $B\to D\pi$ decays.

The decay rate of $\rho^0\pi^\pm$ is sensitive to $\gamma$, while
$\omega\pi^\pm$ is not. For example, ${\cal
B}(B^\pm\to\rho^0\pi^\pm)/ {\cal B}(B^\pm\to\omega\pi^\pm) \sim 1$
for $\gamma\sim 60^\circ$, and ${\cal B}(B^\pm\to\rho^0\pi^\pm)/
{\cal B}(B^\pm\to\omega\pi^\pm)
>1$ for $\gamma>90^\circ$ if $A_0^{B\omega}(0)=A_0^{B\rho}(0)$.
Therefore, a large $\gamma$ preferred by the previous measurement
\cite{Bishai} ${\cal B}(B^\pm\to\rho^0\pi^\pm)=(15\pm5\pm4)\times
10^{-6}$, is no longer favored by the new measurement of
$\rho^0\pi^\pm$.

\section{$B\to\phi K$ Decays}
The previous limit\cite{Bishai} for the branching ratio of
$B^\pm\to\phi K^\pm$ is $0.59\times 10^{-5}$ at 90\% C.L. However,
CLEO has also seen a $3\sigma$ evidence for the decay $B\to\phi
K^*$. Its branching ratio, the average of $\phi K^{*-}$ and $\phi
K^{*0}$ modes, is reported to be \cite{CLEOomega} ${\cal
B}(B\to\phi K^*) =\left(1.1^{+0.6}_{-0.5}\pm 0.2\right)\times
10^{-5}$. Theoretical calculations based on factorization indicate
that the branching ratio of $\phi K$ is similar to that of $\phi
K^*$.  Therefore, it is difficult to understand the
non-observation of $\phi K$.

An observation of the $\phi K$ signal was reported at this
conference to be $(6.4^{+2.5+0.5}_{-2.1-2.0})\times 10^{-6}$ by
CLEO \cite{CLEO} and $(17.2^{+6.7}_{-5.4}\pm1.8)\times 10^{-6}$ by
BELLE \cite{Belle}. The decay amplitude of the penguin-dominated
mode $B\to K\phi$ is governed by $[a_3+a_4+a_5-{1\over
2}(a_7+a_9+a_{10})]$, where $a_3$ and $a_5$ are sensitive to
nonfactorized contributions. In the absence of nonfactorized
effects, we find\cite{CY} ${\cal B}(B^\pm\to\phi
K^\pm)=(6.3-7.3)\times 10^{-6}$, which is in good agreement with
the CLEO result, but smaller than the BELLE measurement.

\section{$B\to K\eta',~K^*\eta$ Decays}
The decays $B\to K^{(*)}\eta(\eta')$ involve interference between
the penguin amplitudes arising from $(\bar uu+\bar dd)$ and $\bar
ss$ components of the $\eta$ or $\eta'$. The branching ratios of
$K\eta'$ ($K^*\eta$) are anticipated to be much greater than
$K\eta$ ($K^*\eta'$) modes owing to the presence of constructive
interference between two comparable penguin amplitudes arising
from non-strange and strange quarks of the $\eta'(\eta)$.

The measured branching ratios of the decays $B\to\eta' K$ are
 \be {\cal B}(B^\pm\to\eta' K^\pm) &=&
\left(80^{+10}_{-~9}\pm 7\right)\times 10^{-6}, \non \\
 {\cal B}(B^0\to\eta' K^0) &=& \left(89^{+18}_{-16}\pm 9
\right)\times 10^{-6},
\en
by CLEO \cite{Richichi} and $(62\pm18\pm8)\times 10^{-6}$,
$<1.12\times 10^{-4}$, respectively by BABAR \cite{BaBar}. The
earlier theoretical predictions in the range of $(1-2)\times
10^{-5}$ are too small compared to experiment. It was realized
later (for a review, see e.g. \cite{CT98}) that $\eta' K$ gets
enhanced because of (i) the small running strange quark mass at
the scale $m_b$, (ii) the sizable $SU(3)$ breaking in the decay
constants $f_8$ and $f_0$, (iii) an enhancement of the form factor
$F_0^{B\eta'}(0)$ due to the smaller mixing $\eta-\eta'$ mixing
angle $-15.4^\circ$ rather than $\approx -20^\circ$, (iv)
contribution from the $\eta'$ charm content, and (v) constructive
interference in tree amplitudes. It was also realized not long ago
that \cite{Kagan} the above-mentioned enhancement is partially
washed out by the anomaly effect in the matrix element of
pseudoscalar densities, an effect overlooked before. As a
consequence, the net enhancement is not very large; we
find\cite{CY} ${\cal B}(B^\pm\to K^\pm\eta')=(40-50)\times
10^{-6}$, which is still smaller than the CLEO result but
consistent with the BELLE measurement. This implies that we
probably need an additional (but not dominant !) SU(3)-singlet
contribution to explain the $B\to K\eta'$ puzzle.

Finally, it is worth remarking that if $\gamma>90^\circ$, the
charged mode $\eta'K^-$ will get enhanced, while the neutral mode
$\eta'K^0$ remains stable \cite{CCTY}. The present data of
$K\eta'$ cannot differentiate between $\cos\gamma>0$ and
$\cos\gamma<0$.

\section{ $B\to\omega K$ and $\rho K$ Decays}
The published CLEO result \cite{CLEOomega} of a large branching
ratio $\left(15^{+7}_{-6}\pm 2\right) \times 10^{-6}$ for
$B^\pm\to\omega K^\pm$ imposes a serious problem to the
generalized factorization approach: The observed rate is
enormously large compared to naive expectation \cite{CCTY}. The
destructive interference between $a_4$ and $a_6$ terms renders the
penguin contribution small. It is thus difficult to understand the
large rate of $\omega K$. Theoretically, it is expected
that\cite{CCTY} ${\cal B}(B^-\to\omega K^-)\gsim 2{\cal
B}(B^-\to\rho^0 K^-)\sim 2\times 10^{-6}$, which now agrees with
the new measurement \cite{CLEO} of ${\cal B}(B^-\to\omega
K^-)<7.9\times 10^{-6}$.


\begin{thebibliography}{99}
\newcommand{\bib}{\bibitem}

\bib{Belle} BELLE Collaboration, P. Chang in these
proceedings.

\bib{BaBar} BABAR Collaboration, T.J. Champion in these
proceedings.

\bib{CLEO} CLEO Collaboration, R. Stroynowski in these
proceedings.

\bib{CCTY} Y.H. Chen {\it et al.,} \pr {\bf
D60}, 094014 (1999).

\bib{Du} D.S. Du {\it et al.,} hep-ph/0005006;
T. Muta {\it et al.,} hep-ph/0006022.

\bib{Stocchi} A. Stocchi in these proceedings.

\bib{Keum} Y.Y. Keum, H.n. Li, and A.I. Sanda, hep-ph/0004173.

\bib{Beneke} M. Beneke {\it et al.,} \prl {\bf 83}, 1914 (1999).

\bib{Beneke1} M. Beneke in these proceedings.

\bib{Cronin} D. Cronin-Hennessy {\it et al.,}
hep-ex/0001010.

\bib{CY1} H.Y. Cheng and K.C. Yang, in preparaion.

\bib{Jessop} C.P. Jessop {\it et al.,}
hep-ex/0006008.

\bib{Bishai} CLEO Collaboration, M. Bishai {\it et al.,} CLEO CONF
99-13, hep-ex/9908018.

\bib{CLEOomega} T. Bergfeld {\it et al.,} \prl
{\bf 81}, 272 (1998).

\bib{CY} H.Y. Cheng and K.C. Yang, \pr {\bf D62}, 054029 (2000).

\bib{Richichi} S.J. Richichi {\it et al.,} \prl {\bf 85}, 520
(2000).

\bib{CT98} H.Y. Cheng and B. Tseng, \pr {\bf D58}, 094005 (1998).

\bib{Kagan} A.L. Kagan and A.A. Petrov, hep-ph/9707354; A. Ali and
C. Greub, \pr {\bf D57}, 2996 (1998).


\end{thebibliography}
\end{document}